\documentclass[aps,prl,twocolumn,superscriptaddress,showpacs,nofootinbib]{revtex4-1}
\usepackage{graphicx}
\usepackage{color}

\begin{document}


\title{
Dipole Polarizability of $^{120}$Sn and Nuclear Energy Density Functionals   
}



\newcommand{\RCNP}{Research Center for Nuclear Physics, Osaka University, Ibaraki, Osaka 567-0047, Japan}
\newcommand{\Wit}{University of the Witwatersrand, Johannesburg 2050, South Africa}
\newcommand{\Kyushu}{Department of Physics, Kyushu University, Fukuoka 812-8581 ,Japan}
\newcommand{\iThemba}{iThemba LABS, Somerset West 7129, South Africa}
\newcommand{\Osaka}{Department of Physics, Osaka University, Toyonaka, Osaka 560-0043, Japan}
\newcommand{\CYRIC}{Cyclotron and Radioisotope Center, Tohoku University, Sendai,  980-8578, Japan}
\newcommand{\CNS}{Center for Nuclear Study, University of Tokyo, Bunkyo,  Tokyo 113-0033, Japan}
\newcommand{\TUDarmstadt}{Institut f\"{u}r Kernphysik, Technische Universit\"{a}t Darmstadt, D-64289 Darmstadt,
Germany}
\newcommand{\Valencia}{Instituto de Fisica Corpuscular, CSIC-Univ. de Valencia, E-46071 Valencia, Spain}
\newcommand{\Gent}{Vakgroep Subatomaire en Stralingsfysica, Univ. Gent, B-9000 Gent, Belgium}
\newcommand{\MSU}{NSCL, Michigan State Univ., MI 48824, USA}
\newcommand{\Kyoto}{Department of Physics, Kyoto University, Kyoto 606-8502, Japan}
\newcommand{\Niigata}{Department of Physics, Niigata University, Niigata 950-2102, Japan}
\newcommand{\RIKEN}{RIKEN Nishina Center, Wako, Saitama 351-0198, Japan}
\newcommand{\HIMAC}{National Institute of Radiological Sciences, Chiba 263-8555, Japan}
\newcommand{\KVI}{Kernfysisch Versneller Instituut, University of Groningen, Zernikelaan 25, NL-9747 AA Groningen,
The Netherlands}
\newcommand{\TexasAM}{Department of Physics and Astronomy, Texas A\&M University-Commerce, Commerce, Texas 75429,
USA}
\newcommand{\GSI}{GSI Helmholtzzentrum f\"{u}r Schwerionenforschung, 64291 Darmstadt, Germany}
\newcommand{\INST}{Institute for Nuclear Science and Technology, 179 Hoang Quoc Viet, Hanoi, Vietnam}
\newcommand{\UEN}{Institut f\"ur Theoretische Physik, Universit\"at Erlangen, D-91054 Erlangen, Germany}
\newcommand{\OsakaU}{Department of Physics, Osaka University, Toyonaka, Osaka, 560-0043, Japan}
\newcommand{\Istanbul}{Physics Department, Faculty of Science, Istanbul University, 34459 Vezneciler, Istanbul,
Turkey}
\newcommand{\Munstar}{Institut f\"{u}r Kernphysik, Westf\"{a}lische Wilhelms-Universt\"{a}t M\"{u}nster, D-48149
M\"{u}nster, Germany}

\author{T. Hashimoto}
\email[E-mail: ]{hasimoto@ibs.re.kr}
\altaffiliation{Present address: Rare Isotope Project, Institute for Basic Science, 70, Yuseong-daero, 1689-gil,
Yuseong-gu, Daejeon, Korea}
\affiliation{\RCNP}
\author{A.~M.~Krumbholz}\affiliation{\TUDarmstadt}
\author{P.-G. Reinhard}\affiliation{\UEN}
\author{A.~Tamii}\affiliation{\RCNP}
\author{P.~von Neumann-Cosel}\email[E-mail: ]{vnc@ikp.tu-darmstadt.de}\affiliation{\TUDarmstadt}
\author{T.~Adachi}\affiliation{\OsakaU}
\author{N.~Aoi}\affiliation{\RCNP}
\author{C.~A.~Bertulani}\affiliation{\TexasAM}
\author{H.~Fujita}\affiliation{\RCNP}
\author{Y.~Fujita}\affiliation{\RCNP}
\author{E. Ganio\v{g}lu}\affiliation{\Istanbul}
\author{K.~Hatanaka}\affiliation{\RCNP}
\author{E.~Ideguchi}\affiliation{\RCNP}
\author{C.~Iwamoto}\affiliation{\RCNP}
\author{T.~Kawabata}\affiliation{\Kyoto}
\author{N.~T.~Khai}\affiliation{\INST}
\author{A.~Krugmann}\affiliation{\TUDarmstadt}
\author{D.~Martin}\affiliation{\TUDarmstadt}
\author{H.~Matsubara}\affiliation{\HIMAC}
\author{K.~Miki}\affiliation{\RCNP}
\author{R.~Neveling}\affiliation{\iThemba}
\author{H.~Okamura}\affiliation{\RCNP}
\author{H.~J.~Ong}\affiliation{\RCNP}
\author{I.~Poltoratska}\affiliation{\TUDarmstadt}
\author{V.~Yu.~Ponomarev}\affiliation{\TUDarmstadt}
\author{A.~Richter}\affiliation{\TUDarmstadt}
\author{H.~Sakaguchi}\affiliation{\RCNP}
\author{Y.~Shimbara}\affiliation{\CYRIC}
\author{Y.~Shimizu}\affiliation{\RIKEN}
\author{J.~Simonis}\affiliation{\TUDarmstadt}
\author{F.~D.~Smit}\affiliation{\iThemba}
\author{G.~S\"{u}soy}\affiliation{\Istanbul}
\author{T.~Suzuki}\affiliation{\RCNP}
\author{J.~H.~Thies}\affiliation{\Munstar}
\author{M.~Yosoi}\affiliation{\RCNP}
\author{J.~Zenihiro}\affiliation{\RIKEN}


\date{\today}

\begin{abstract}
The electric dipole strength distribution in $^{120}$Sn between 5 and 22 MeV has been determined at RCNP Osaka from polarization transfer observables measured in proton inelastic scattering at $E_0 =295$ MeV and forward angles including $0^\circ$.
Combined with photoabsorption data a highly precise electric dipole polarizability $\alpha_\mathrm{D}(^{120}{\rm Sn}) = 8.93(36)$~fm$^3$ is extracted.
The dipole polarizability as isovector observable par excellence carries direct information on the nuclear symmetry energy and its density dependence.
The correlation of the new value with the well established $\alpha_\mathrm{D}(^{208}{\rm Pb})$ 
serves as a test of its prediction by nuclear energy density functionals (EDFs).
Models based on modern Skyrme interactions describe the
data fairly well while most calculations based on relativistic Hamiltonians cannot.
\end{abstract}


\maketitle


 
The nuclear equation of state (EOS) describing the energy of nuclear
matter as function of its density has wide impact on nuclear physics
and astrophysics \cite{hor14} as well as physics beyond the standard
model~\cite{wen09,pol99}.  The EOS of symmetric nuclear matter with
equal proton and neutron densities is well constrained from the ground
state properties of finite nuclei, especially in the region of
saturation density $\rho_{0} \simeq 0.16$ fm$^{-3}$ \cite{dan02}.
However, the description of astrophysical systems as, e.g., neutron
stars requires knowledge of the EoS for asymmetric matter
\cite{lat04,Sto06,lat14,heb10} which is related to the leading
isovector parameters of nuclear matter, viz.\ the symmetry energy ($J$)
and its derivative with respect to density ($L$) \cite{mol95ra}.  For a
recent overview of experimental and theoretical studies of the
symmetry energy see Ref.~\cite{epj50}. In spite of steady
  extension of knowledge on exotic nuclei, just these isovector
properties are poorly determined by fits to experimental ground state
data because the valley of nuclear stability is still extremely
narrow along isotopic chains \cite{klu09,Erl13,naz14}. Thus one needs
observables in finite nuclei specifically sensitive to
isovector properties to better confine $J$ and $L$.  There are two such
observables, the neutron skin $r_\mathrm{skin}$ in nuclei with large
neutron excess and the (static) dipole polarizability
$\alpha_\mathrm{D}$.

The neutron skin thickness
$r_\mathrm{skin}=\langle{r}\rangle_n-\langle{r}\rangle_p$ defined as
the difference of the neutron and proton root-mean-square radii 
$\langle{r}\rangle_{n,p}$ is determined by the interplay between the
surface tension and the pressure of excess neutrons on the core described by $L$ \cite{bro00,fur02}.  Studies within 
nuclear density-funtional theory \cite{Ben03}
show for all EDFs
a strong correlation between $r_\mathrm{skin}$ and the
isovector symmetry energy parameters \cite{roc11,pie12,erl14}.  The most
studied case so far is $^{208}$Pb, where $r_\mathrm{skin}$ has been
derived from coherent photoproduction of $\pi^{0}$ mesons
\cite{tar14}, antiproton annihilation \cite{klo07,bro07}, proton
elastic scattering at 650 MeV \cite{sta94} and 295~MeV \cite{zen10},
and from the dipole polarizability \cite{tam11}.  A nearly
model-independent determination of the neutron skin is possible by
measuring the weak form factor of nuclei with parity-violating elastic
electron scattering \cite{hor01}.  Such an experiment has been
performed for $^{208}$Pb but the statistical uncertainties are still
too large for serious constraints of the neutron skin \cite{abr12}.

A particularly useful experimental observable to constrain the large theoretical uncertainties on $J$ and $L$ is $\alpha_\mathrm{D}$ \cite{rei10} which can determined by a weighted integral over the photoabsorption cross section $\sigma_\mathrm{abs}$ \cite{boh81}
\begin{equation}
\label{eq:pol}
  \alpha_\mathrm{D}
  =
  \frac{\hbar c}{2\pi^{2} } 
  \int \frac{\sigma_\mathrm{abs}}{E_{\rm x}^{2}}{\rm d}E_{\rm x} 
  = 
  \frac{8 \pi}{9} \int\frac{{\rm d}{B{\rm (E1)}}}{E_{\rm x}}{\rm d}E_{\rm x},
\end{equation}
where $E_{\rm x}$ is the excitation energy and $B$(E1) the reduced electric dipole transition strength.  
It is the aim of this letter to present a new experimental result for $\alpha_\mathrm{D}$ in a
heavy nucleus, $^{120}$Sn.  This data point is then used
together with the well established
$\alpha_\mathrm{D}(^{208}\mathrm{Pb})$ to scrutinize EDFs.

The $E1$ response is dominated by excitation of the isovector giant
dipole resonance (IVGDR) well known in many nuclei from
photoabsorption experiments.  Because of the inverse energy weighting
in Eq.~(\ref{eq:pol}), $\alpha_\mathrm{D}$ also depends on the
low-energy strength studied mainly with the $(\gamma,\gamma^\prime)$
reaction.  However, extraction of the $E1$ strength from
$(\gamma,\gamma^\prime)$ data is rather model-dependent \cite{sav13}.

Recently, polarized inelastic proton scattering at 295 MeV and at forward angles including $0^\circ$ has been established as a new method to extract the complete $E1$ strength in heavy nuclei from low excitation energy across the giant resonance region \cite{tam11}.  
In this particular kinematics selective excitation of $E1$ and spin-$M1$ dipole modes is observed.  
Their contributions to the cross sections can be separated either by a multipole decomposition analysis (MDA) \cite{pol12} or independently by measurement of a combination of polarization transfer observables (PTA) \cite{tam11}.  Good agreement of both methods was demonstrated for the reference case $^{208}$Pb where values of $r_\mathrm{skin}$ and $L$ derived from $\alpha_\mathrm{D}(^{208}\mathrm{Pb})$ conform with results from other methods \cite{tam14}.
 
All EDFs agree on showing strong correlations between
$\alpha_\mathrm{D}$, $r_\mathrm{skin}$, $J$, and $L$, but the actual
predictions of $\alpha_\mathrm{D}$ for given $J$ and $L$ values differ
considerably.  While the result for $^{208}$Pb \cite{tam11} already
excluded many older Skyrme interactions, modern Skyrme-Hartree-Fock
(SHF) and relativistic mean-field (RMF) models can be brought into
agreement e.g.\ by changing $J$, which can be varied over a certain
range without deteriorating the fit of the interaction
parameters \cite{klu09}.  Experimental information on
$\alpha_\mathrm{D}$ in other nuclei is therefore essential to
further constrain the isovector part of the EDF interaction.
We note that some information on E1 strength distributions in heavy
neutron-rich nuclei is available \cite{adr05,kli07,wie09,ros13} but an
extraction of $\alpha_{\rm D}$ from these results is completely model-dependent, in
contrast to the data discussed here.
   
Here, we report on a measurement of the electric dipole response in
$^{120}$Sn with polarized proton scattering based on a PTA covering
excitation energies $5-22$ MeV.  $E1$ strength in $^{120}$Sn below 5
MeV was measured by ($\gamma,\gamma'$)~\cite{oze14} and above neutron
threshold by ($\gamma,xn$) \cite{ful69,lep74,uts11} experiments.  A
combination of all available data enables a precise determination of
$\alpha_\mathrm{D}$.  The E1 strength has also been determined from a
MDA of the $(p,p^\prime)$ cross sections \cite{kru15} but
photoabsorption cross sections had to be included as constraints and
therefore the result - in contrast to the PTA - is not independent
from these data.

The experiment was performed at the RING cyclotron facility of the
Research Center for Nuclear Physics (RCNP), Osaka University, Japan.
Details of the experimental technique can be found in
Ref.~\cite{tam09}.  A polarized proton beam was accelerated to
$E_0=295$~MeV and scattered protons were momentum-analyzed with the
Grand Raiden spectrometer \cite{fuj99} placed at $0^\circ$ covering an
angular and excitation energy range of $0^\circ -2.5^\circ$ and $5 -
22$ MeV, respectively.  An isotopically enriched (98.4\%)
self-supporting $^{120}$Sn foil with a thickness of 6.5 mg/cm$^2$
served as a target.  The beam intensity was $1-2$~nA with an average
polarization of 0.7.

A decomposition of spinflip and non-spinflip cross sections can be achieved \cite{suz00} by the combined information of the polarization transfer observables $D_{LL}$, $D_{SS}$ and $D_{NN}$ \cite{ohl72} determined in a secondary scattering experiment.
Since $D_{SS}$ and $D_{NN}$ are indistinguishable at $0^\circ$, only $D_{LL}$ and $D_{SS}$ were measured in the present experiment.
It is convenient to introduce the total spin transfer
\begin{equation}
\label{eq:sigma}
\Sigma = \frac{3-2D_{SS}-D_{LL}}{4},
\end{equation}
which takes values of zero for non-spinflip and one for spinflip transitions.
Because of the different reaction mechanism these can be identified with $E1$ (Coulomb excitation) or $M1$ (spin-isospinflip part of the proton-nucleus interaction) excitations, respectively.

\begin{figure}[tb]
\includegraphics[width=20pc]{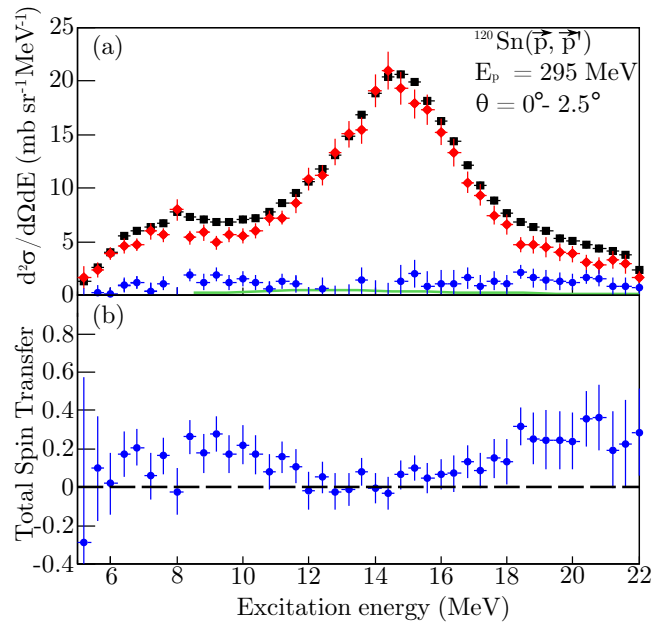}
\caption{\label{fig1}
(Color online). 
(a) Double differential cross sections (black) and decomposition into non-spinflip (red) and spinflip (blue) parts of
the $^{120}$Sn($\vec{p},\vec{p}^\prime$) reaction at $E_0 = 295$ MeV and $\theta = 0^\circ - 2.5^\circ$.
The green line shows the cross sections due to excitation of the ISGQR estimated as described in the text.
(b) Total spin transfer from Eq.~(\ref{eq:sigma}).
}
\end{figure}
Figure~\ref{fig1}(a) displays the measured cross sections (black circles)  in 400 keV bins.
The bump structure centered at E$_{x} \simeq 15$ MeV corresponds to the IVGDR.
The extracted total spin transfer [Fig.~\ref{fig1}(b)] is almost zero in this energy region as expected for Coulomb excitation and approaches maximum values of about 0.2 around 9 MeV (the location of the $M1$ spinflip resonance \cite{hey10}) and above 18 MeV.
The decomposition into non-spinflip and spinflip parts is shown in Fig~\ref{fig1}(a) by red and blue circles, respectively.
The non-spin-flip cross sections contain an small $E2$ contribution (green line) from nuclear excitation of the isoscalar giant quadrupole resonance (ISGQR).
It was determined using the isoscalar $B$(E2) strength distribution~\cite{li10} as described in Ref.~\cite{kru15} and never exceeds 4\% in a single bin.
%
%

\begin{figure}[tbh]
\includegraphics[width=20pc]{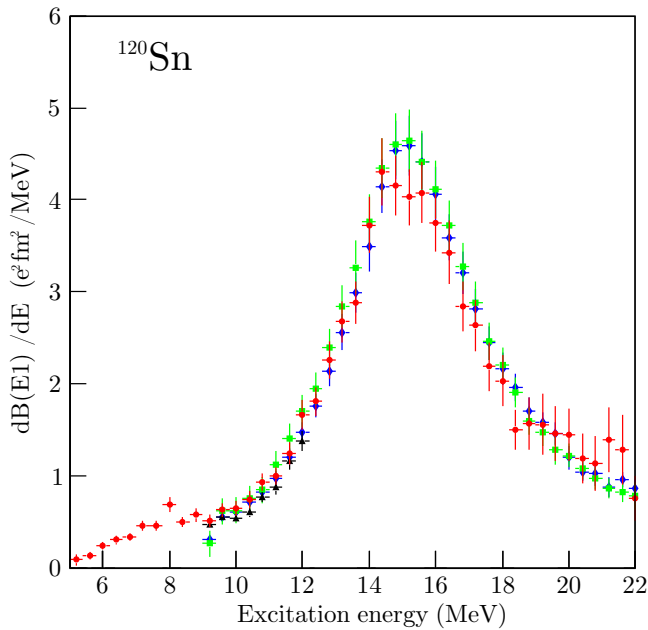}
\caption{\label{fig2}
(Color online).
Comparison of the $B$(E1) strength distribution in $^{120}$Sn determined by the present work (red circles) and in 
($\gamma,xn$) experiments (blue circles \cite{ful69}, green circles~\cite{lep74}, and black circles~\cite{uts11}).
}
\end{figure}
Figure~\ref{fig2} shows the $B$(E1) strength distribution (red
circles) deduced from the $\Delta S = 0$ cross sections assuming
semiclassical Coulomb excitation \cite{ber88}.  The photoabsorption
data converted to $B$(E1) strength are shown as blue \cite{ful69},
green \cite{lep74}, and black \cite{uts11} circles, respectively.  All
data agree well with each other.  Near the IVGDR maximum slightly
smaller values are found in the present work but they still accord
within the experimental uncertainties.  Since the MDA analysis below
neutron threshold does not depend on the photoabsorption data, the
$B$(E1) strength up to 9 MeV can be compared to the present result.
Good agreement of the summed strengths from MDA [48.7(29) e$^2$fm$^2$] and PTA [54.1(41) e$^2$fm$^2$] 
and the corresponding contribution to $\alpha_\mathrm{D}$ is
observed.

We now turn to the determination of the electric dipole polarizability
in $^{120}$Sn taking into account all available data.  
The energy region below 5 MeV makes a negligible ($< 0.1$\%) 
contribution to $\alpha_\mathrm{D}$ \cite{oze14}.  Results for
the energy region from 5 to 10 MeV are taken from the present work
\cite{pad} and amount to 1.12(7) fm$^3$, contributing about 12.5\% to
the total value.  The main contribution, 7.00(29) fm$^3$, stems from
the IVGDR region, where the present results and those from
Refs.~\cite{ful69,lep74,uts11} were averaged between 10 and 22 MeV.
Between 22 and 28.9~MeV data are available from Ref.~\cite{ful69},
0.51(6) fm$^3$.  Finally, the polarizability at even higher energies
up to 135 MeV was taken from a $^{\rm nat}$Sn($\gamma,xn$) experiment
\cite{lep81} neglecting an isotopic dependence.  The contribution,
0.31(10) fm$^3$, is small but non-negligible considering the final
precision achieved.  In total, we find $\alpha_\mathrm{D}(^{120}{\rm
  Sn}) = 8.93(36)$ fm$^3$, where the error contains the statistical
and systematic uncertainties of all data used.

\begin{figure}[tbhp]
\includegraphics[width=15pc]{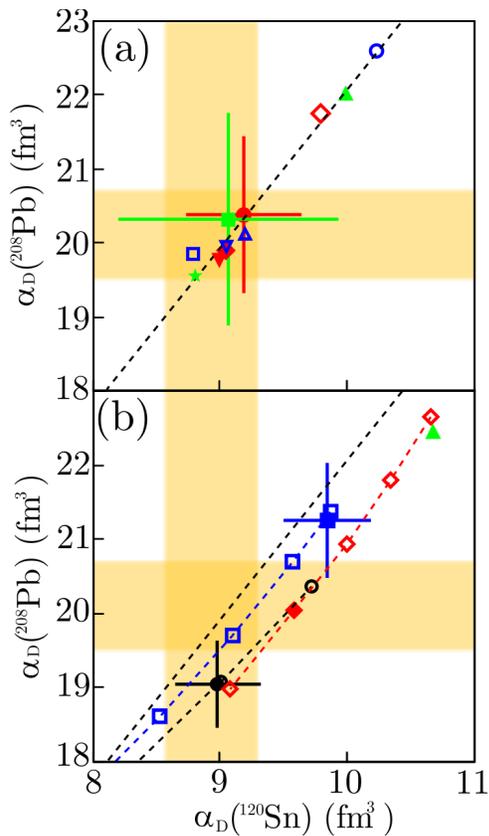}
\caption{\label{fig3} (Color online).  Correlation of the experimental
  $\alpha_\mathrm{D}$ values for $^{120}$Sn and $^{208}$Pb with
  uncertainties shown as yellow bands.  (a) Comparison with
  models based on Skyrme interactions SkM$^*$~\cite{bar82} (red
  square), SkP~\cite{dob84} (blue triangle), SkT6~\cite{ton84} (red
  diamond), SG-II~\cite{gia81} (blue circle), SkI3~\cite{rei95} (green
  triangle), SLy6~\cite{cha97} (red triangle), BSk4~\cite{gor03}
  (green circle), SV-bas (blue diamond) \cite{klu09} and
  UNEDF2~\cite{kor14} (blue square).  The SV-min~\cite{klu09} (red
  circle) and RD-min~\cite{erl10} (green square) interactions
  additionally provide theoretical error bars \cite{klu09,dob14}.
  The dashed black line indicates the correlation between both
    $\alpha_\mathrm{D}$ values.  
(b) Comparison with relativistic mean field models DD-PC-min (blue squares) and DD-ME-min (black circles), both  from Ref.~\cite{naz14}, FSU~\cite{tod05} (red diamonds), and FSU2 \cite{che14} (green triangle).  Full symbols denote the results of optimum parameter sets.  Open symbols show results varying the symmetry energy parameter {\it J}.  The  dashed lines serve to guide the eye.  The dashed black line from (a) is repeated for direct comparison of SHF and RMF models.
}
\end{figure}
Having now at hand precise data for
$\alpha_\mathrm{D}(^{120}\mathrm{Sn})$ and
$\alpha_\mathrm{D}(^{208}\mathrm{Pb})$, we use them to scrutinize the
performance of a broad variety of EDFs from SHF and RMF, all values
including pairing at the BCS level.  The theoretical
$\alpha_\mathrm{D}$ values are computed from the static response to an external
dipole field. Figure~\ref{fig3} displays the EDF results for
$\alpha_\mathrm{D}(^{208}\mathrm{Pb})$ versus
$\alpha_\mathrm{D}(^{120}\mathrm{Sn})$ together with the experimental
values indicated by yellow bands.  

Panel (a) collects SHF results for
a couple of widely used parametrizations (see caption).  Although
taken from very different sources, all SHF results together show
  a strong correlation between the theoretical $\alpha_\mathrm{D}$ values
  indicated by the dashed black line. The actual position on the line
  is determined by $J$ and the large span of results along the line
demonstrates the uncertainty in $J$.  Note, however, that the majority
of SHF results resides nicely within the experimentally allowed yellow
square. The outliers are all rather old parametrizations adjusted
before appearance of the many data on neutron rich nuclei.  The fact
that the linear trend goes right through the experimentally allowed
square and that most parametrization lies within indicates that the
isovector density dependence of SHF is realistic.  

Two parametrizations (SV-min, RD-min) are shown together with error bars from statistical
analysis \cite{klu09,dob14}. These are larger than the
experimental uncertainties demonstrating that the data provide indeed useful constraints on the isovector
parameters \cite{naz14}.

Simple error bars hide the linear correlation discussed above.  
This can be better visualized by series of
parametrizations with systematically varied $J$
\cite{klu09,naz14,erl14}.  We do this in connection with RMF approaches shown in
panel (b).  Unlike SHF, there is greater variance in modeling density
dependence for RMF.  We consider three variants thereof: the density
dependent point-coupling model (DD-PC) \cite{zha10}, the density
dependent meson-exchange model (DD-ME) \cite{lal05}, and non-linear
meson coupling in FSU \cite{tod05}.  For all three cases we show
series with varied $J$ (open symbols) and the best fit (full symbols with error bars where available).
For better comparability, the
  series DD-ME, DD-PC, and RD were fitted to the same data pool as SV-min and RD-min
  \cite{klu09}. 
(Fit procedures for the FSU family are described in Refs.~\cite{tod05,che14}).
  However, the actual fit strategy seems to be of lower importance as
  the original standard parametrizations DD-ME2 \cite{lal05} and PC-1
  \cite{zha10} lie again on the corresponding lines in the plot. 
All sets are strongly correlated with nearly linear trends, however, with
different offset depending on the form of the EDF.  While the SHF series
goes approximately through the center of the experimental square, all RMF
chains are off center, two of them just touching the square.
Only DD-PC comes closer and only one DD-PC parametrization (with
$J=32$ MeV) lies within the correlation box.  The best-fit
parametrizations (full symbols) are all outside.
This indicates that RMF models still need to be improved in
  the isovector channel \cite{naz14}, although the modern,
density-dependent functionals already constitute large progress in this
respect in comparison to older RMF functionals \cite{rei89}.

\begin{figure}[tbh]
\includegraphics[width=20pc]{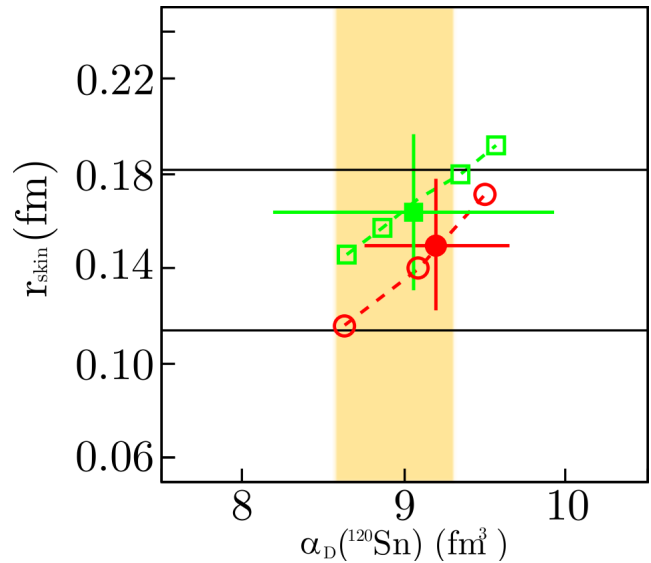}
\caption{\label{fig4}
(Color online).
Relationship between $\alpha_\mathrm{D}$ and $r_\mathrm{skin}$ for $^{120}$Sn predicted by the SV-min~\cite{klu09} (red
circle) and RD-min~\cite{erl10} (green square) interactions.
Full symbols are the results of the optimum parameter sets and open symbols correspond to a variation of the symmetry
energy parameter $J$ as in Fig.~\ref{fig3}(b).
Dashed lines are to guide the eye.
The horizontal lines denote the range of $r_\mathrm{skin}$ values compatible with the experimental polarizability shown
as yellow band.
}
\end{figure}
Using the strong correlation between $\alpha_\mathrm{D}$ and $r_\mathrm{skin}$ \cite{rei10} one can derive the neutron skin thickness of $^{120}$Sn from EDFs capable to describe the data in Fig.~\ref{fig3}.  
A similar analysis has been performed for $^{208}$Pb \cite{pie12}.  
Since the models are not independent, rather than averaging (as done in Ref.~\cite{pie12}) we take the SV-min (red circles) and RD-min (green squares) results as representative and estimate the theoretical uncertainties.
Figure~\ref{fig4} shows the predictions of the correlation between $r_\mathrm{skin}$ and $\alpha_\mathrm{D}$.  As for the relativistic models, a variation of $J$ (open symbols) is compatible with the optimum fits (full symbols).  
The range of values consistent with the experimental polarizability indicated by the horizontal lines corresponds to $r_\mathrm{skin} = 0.148(34)$ fm.  
The result is in good agreement with values extracted from measurements of the spin-dipole resonance \cite{kra99}, 0.18(7) fm, and proton elastic scattering \cite{ter08}, 0.16(3) fm, while antiproton annihilation \cite{sch03} finds a much smaller value, 0.08(+3)(-4) fm.
 
In summary, we have measured polarized proton inelastic scattering off $^{120}$Sn at very forward angles and extracted the $E1$ strength distribution between 5 and 22 MeV by an analysis of polarization transfer observables.
Combining the present results with $(\gamma,xn)$ data, the dipole polarizability could be extracted with a precision of 4\%.
The correlation with the polarizability of $^{208}$Pb \cite{tam11} provides an important test of EDFs indispensable for the extraction of properties of the symmetry energy in neutron-rich matter.
While modern Skyrme interactions can describe the data, in contrast to most RMF calculations.
With the typical theoretical uncertainties indicated, the combined data from $^{208}$Pb and $^{120}$Sn provide an important constraint to improve the description of static isovector properties in EDFs.

Considering the importance of polarizability data, a systematic study at different shell closures and exploration of the role of deformation is called for.
One important future project is a systematic measurement of $\alpha_\mathrm{D}$ covering the range of stable tin isotopes \cite{vnc14}.
Together with a new measurement of relativistic Coulomb excitation of the neutron-rich tin isotopes $^{124-134}$Sn at GSI \cite{aum12} a unique set of data will be available to investigate the impact of neutron excess on the formation of a neutron skin in a set of nuclei with similar underlying structure.

We thank the RCNP accelerator staff for excellent beams. 
J.~Piekarewicz kindly provided us with the FSU and FSU2 results shown in Fig.~\ref{fig3}(b).
This work was supported by JSPS (Grant No. 25105509), DFG (contracts SFB 634 and NE 679/3-1), and BMBF (contract 05P12RFFTG).
N.T. Khai acknowledges support from NAFOSTED of Vietnam under grant 103.01-2011.17.



\begin{thebibliography}{00}
\bibitem{pad}
%
Photoabsorption data between $S_n$ and 10 MeV are not considered because of potentially large systematic
uncertainties close to threshold.
\end{thebibliography}
\end{document}